\documentclass[letterpaper]{JHEP3}
\usepackage{epsfig}
\newcommand{\beq}{\begin{equation}}
\newcommand{\eeq}{\end{equation}}
\newcommand{\bea}{\begin{eqnarray}}
\newcommand{\eea}{\end{eqnarray}}
\preprint{{\tt hep-th/0305059}}

\author{ Fr\'ed\'eric Leblond\footnote{E-mail: {\tt
fleblond@perimeterinstitute.ca}} \ and Amanda W.
Peet\footnote{E-mail: {\tt peet@physics.utoronto.ca}}\\ $^{*}$
Department of Physics, McGill University, Montr\'eal, Qu\'ebec H3A
2T8 Canada \\ $^{*}$ Perimeter Institute for Theoretical Physics,
Waterloo, Ontario N2J 2W9 Canada \\ $^{\dagger}$ Department of
Physics,
University of Toronto, Toronto, Ontario M5S 1A7 Canada \\
$^{\dagger}$ Cosmology and Gravity Program, Canadian Institute for
Advanced Research}

\date{May, 2003}

\title{A note on the singularity theorem for supergravity SD-branes}

\abstract{Recently, a singularity theorem for full SD-brane
spacetimes was given in hep-th/0305055. We comment on the relation
between this and previous work as well as provide a more geometric
formulation interpreted as a no-go theorem. We then point out that
some setups of physical interest escape the theorem: cosmological
applications, half-SD$p$-branes and decaying unstable D$p$-branes
for general $p$. We also provide indications that the
space-filling full SD8-brane (in $d=10$) escapes as well, because
of the important r{\^{o}}le of Ramond-Ramond fields.  In any case,
tachyon cosmology is not ruled out by the no-go theorem.  Lastly,
we remark upon interesting directions for potential
generalizations of the theorem, and quantum corrections.}


\begin{document}

\baselineskip16pt
%\parskip=4pt

%--------------------------------------------------------------------+
\section{Introduction}

SD-branes, introduced in \cite{andy1}, are objects of considerable
physical interest.  There have been a number of works elucidating
their physical properties.  Some investigations were done using CFT
methods, {\it i.e.}, including stringy $\alpha^\prime$ corrections but
no $g_s$ loop effects or gravitational backreaction
\cite{andythermo,stro2,stro1}.  Others were focused on the problem of
finding the supergravity fields exerted by a large number of
coincident SD-branes \cite{gut,kmp,blw,sbranes,us2}.

SD-branes are not BPS, and they are not stable either.  It is not {\em
a priori} clear whether there should exist a totally smooth
supergravity description of these objects, {\it i.e.}, solutions
without singularities. Several groups have investigated this question
at various levels of approximation.  Initial works included just the
bulk supergravity fields in the analysis \cite{gut,kmp}. Later works
\cite{blw,us2} concentrated on the r{\^{o}}le of the homogeneous
tachyon, the most relevant open-string degree of freedom in the
problem. The hope was that including the backreaction between the
rolling tachyon and the bulk fields might resolve the singularities
previously found.  In this analysis, the tachyon starts at or near the
top of its potential hill, and subsequently rolls down the hill
sending its energy into bulk supergravity modes.

One of those recent papers \cite{us2}, by the current set of
authors, exhibited numerical results indicating singularity-free
solutions. This was however not the definitive word on the
subject, for several reasons, one being the inability to handle a
sufficiently general ansatz for the bulk fields and tachyon.
Another reason was a numerical code instability.  One source of
numerical instability was understood and discussed - it happened
even for the tachyon in flat space.  However, it turns out that
there was another more important source which was not recognized
at that time.

Most recently, a very interesting singularity theorem for full
SD-branes was given in \cite{buchelwalcher}. This
work\footnote{All equations are in Einstein frame} showed {\em
analytically} that {\em full} SD$p$-branes with codimension one or
greater {\em must} develop singularities either in the past or the
future of $t=0$. This theorem applies to solutions of the type
studied in \cite{us2}, and provides significant progress in the
general problem of finding SD-brane supergravity solutions.

We begin this note by briefly reviewing the singularity theorem
itself, and giving an alternate formulation.  Using it, we show
precisely where the numerical code of \cite{us2} went astray.  The
upshot is that the singularity occurred in the past of $t=0$. This
is an inescapable conclusion, given the fact that at least one
Hubble parameter in the problem {\em must} have a nonzero (small)
derivative at $t=0$ in order to satisfy the constraint equation.
We then move to showing two situations of notable interest where
the no-go theorem does {\em not} hold: cosmological applications
and decaying unstable branes. We also give strong indications that
space-filling SD8-branes may escape the no-go theorem completely.
In the discussion, we remark on what may resolve the
singularities. We also point out several directions in which the
no-go theorem could usefully be generalized.  One of them is
relaxing the restrictiveness of the ansatz for the bulk and
tachyon fields; another is allowing tachyon inhomogeneity.

%--------------------------------------------------------------------+
\section{The no-go theorem for full SD-branes}
\label{nogot}

There can be two types of open string tachyon evolutions that could
correspond to full-SD-branes: \begin{itemize} \item The tachyon evolves
from one side of the potential $V(T)$ ($T\rightarrow \pm\infty$) up
and over to the other side corresponding to $T\rightarrow \mp \infty$;
\item the tachyon evolves from one side of the potential
($T\rightarrow \pm \infty$) but never reaches the maximum at
$T=0$. There is then a turning point ($\dot{T}=0$) for some
$|T|>0$ and the future evolution of the SD-brane proceeds towards
$T\rightarrow \pm \infty$.
\end{itemize}
The singularity theorem of \cite{buchelwalcher} clearly states
that the corresponding tachyon evolutions will {\it always} lead
to a curvature singularity either in the past (big-bang) or the
future (big-crunch) depending on the boundary conditions on the
tachyon and the supergravity fields. This theorem assumes an
homogeneous tachyon field, $T=T(t)$, and a metric ansatz of the
form
\beq ds^2 = -dt^2 + a^2_\parallel(t) dx_\parallel^2 + a^2_\perp(t)
dx_\perp^2, \eeq where $dx_\parallel^2$ and $dx_\perp^2$ are
maximally symmetric spaces of flat ($k_{\parallel}=0$ or
$k_{\perp}=0$), positive ($k_{\parallel}=1$ or $k_{\perp}=1$) or
negative ($k_{\parallel}=-1$ or $k_{\perp}=-1$) curvature.
Each constant time Cauchy surface of the SD-brane is associated
with the volume function \beq \label{volume} V_{{\rm S}} = v \,
a_{\parallel}^{p+1} a_{\perp}^{8-p} = v \, a^{9}, \eeq where $v$
is a constant. The systems under consideration are either Type IIa
or Type IIb supergravity. As described in \cite{blw,us2} the setup
consists in an unstable brane source (the instability being driven
by the corresponding open string tachyon) coupled to the massless
supergravity fields: graviton, dilaton and Ramond-Ramond field.
The ansatz for the dilaton and the Ramond-Ramond field strength
are respectively \beq \Phi=\Phi(t), \;\;\; {\rm and} \;\;\;
F_{tx_{1}...x_{p+1}} = A(t) a_{\parallel}^{p+1}. \eeq The theorem
in \cite{buchelwalcher} relies on the following combination of the
Einstein equations, \beq \label{bcond} 9\frac{\ddot{a}}{a} = -
\left[ I_{{\rm sugra}} + I_{{\rm tachyon}} \right], \eeq where
\beq\label{Isugra} I_{{\rm sugra}} = {\frac{1}{4}} \left[
2\dot{\Phi}^2 + {\frac{(7-p)}{4}} e^{\Phi(3-p)/2} A^2 \right] +
\frac{(p+1)(8-p)}{9}\left(\frac{\dot{a_{\parallel}}}
{a_{\parallel}}-\frac{\dot{a_{\perp}}}{a_{\perp}} \right)^{2},
\eeq $A(t)$ is an ingredient in the Ramond-Ramond flux, and
\beq \label{Itachyon2} I_{{\rm tachyon}} = \frac{\lambda V(T)
e^{\Phi\left(\frac{p}{4}-\frac{1}{2}\right)}}{16 a_{\perp}^{8-p}}
\left( \frac{7}{\sqrt{\Delta}} -(p+1)\sqrt{\Delta} \right), \eeq
with $\Delta \equiv 1 - e^{-\Phi/2}{\dot{T}}^2$ and $\lambda$ is a
constant quantifying the backreaction between the rolling tachyon
and bulk fields
(see \cite{us2} for a definition). The quantity $\Delta$ starts out at
(or near) unity at $t=0$, and becomes monotonically smaller with time.

The singularity theorem of \cite{buchelwalcher} states that for
full-SD-brane evolution a singularity will develop unless the
gravitational sources introduced provide enough leeway that
$\ddot{a}/a$ can become positive. Clearly, $I_{{\rm sugra}}\geq 0$
so in the context of pure supergravity (no explicit introduction
of extended sources) the volume of a full-SD-brane will never
experience a period of positive acceleration. This should be
related with failed earlier attempts at finding non-singular
SD-brane solutions in the context of supergravity
\cite{andy1,kmp,gut}. The hope was \cite{blw,us2} that the tachyon
might be able to resolve this singularity. It can be easily shown
that for $p<7$ we have $I_{{\rm tachyon}}\geq 0$ which implies
that introducing a tachyon source cannot be used to circumvent the
singularity theorem.

The most general supergravity solutions obtained in \cite{kmp}
were associated with a worldvolume which is anisotropic while the
ansatz used to derive the singularity theorem in
\cite{buchelwalcher} is isotropic. Only in the anisotropic case
were there solutions free of curvature singularities. Presumably
the anisotropy modifies the RHS of eq.~(\ref{bcond}) allowing the
volume to gain positive acceleration during its evolution in such
a way as to avoid big-crunch or big-bang singularities.
%%%
Also, tachyon inhomogeneity, and any effect on the bulk fields'
ansatz, was not handled in the singularity theorem either.  Work
on this and other inhomogeneity questions is in progress
\cite{inhomogtree}.

However, for $p=8$, $I_{{\rm tachyon}}$ can be
negative\footnote{For $p=7$ $I_{{\rm tachyon}}$ can also be
negative but \cite{buchelwalcher} showed that there is
nevertheless a singularity theorem in this case.} which means the
tachyon can drive the volume of the SD-brane into periods of
positive acceleration therefore potentially avoiding the
big-crunch or big-bang singularities predicted by the theorem.  We
provide some remarks on this specific case in the next section.

In short, the singularity theorem implies that full-SD-branes are
singular in one of two ways. Type I: full-SD-branes will evolve
out of a big-bang singularity (the singularity is in the past of
the tachyon evolution) or, Type II: full SD-branes lead to a
big-crunch singularity in the future. The results of
\cite{buchelwalcher} clearly show that during the evolution of the
tachyon the Hubble function \beq H = \frac{(p+1)}{9} H_\parallel +
\frac{(8-p)}{9} H_\perp , \eeq
where $H_{\parallel}= \dot{a}_{\parallel}/a$ and $H_{\perp}=
\dot{a}_{\perp}/a$, will diverge in finite time. Inspection of the
associated Ricci scalar expression
%%%
\beq\label{ricciscalar} \begin{array}{rl}  {\cal R}(t) = &
-2(p+1)(8-p)H_{\parallel}H_{\perp} -p(p+1)
{\displaystyle{ \left[ H_{\parallel}^{2}
+ \frac{k_{\parallel}}{a_{\parallel}^{2}} \right]
-(8-p)(7-p) \left[
H_{\perp}^{2} + \frac{k_{\perp}}{a_{\perp}^{2}} \right] }}
\cr &
+\frac{1}{4}\left[ (p+1)(p-7)P_{\parallel} -(8-p)P_{\perp} +
8(p+1)\rho \right] \end{array} \eeq
%%%
shows that, unless there is an unlikely conspiracy among terms,
this corresponds to a curvature singularity.  In the above
expression, $\rho$ is the energy density, $P_\parallel$ the
pressure in the unstable brane worldvolume, and $P_\perp$ the
pressure in the transverse directions.  (Details are in
\cite{buchelwalcher} Table 1 and will be shown here only as
needed.)

\FIGURE{\epsfig{file=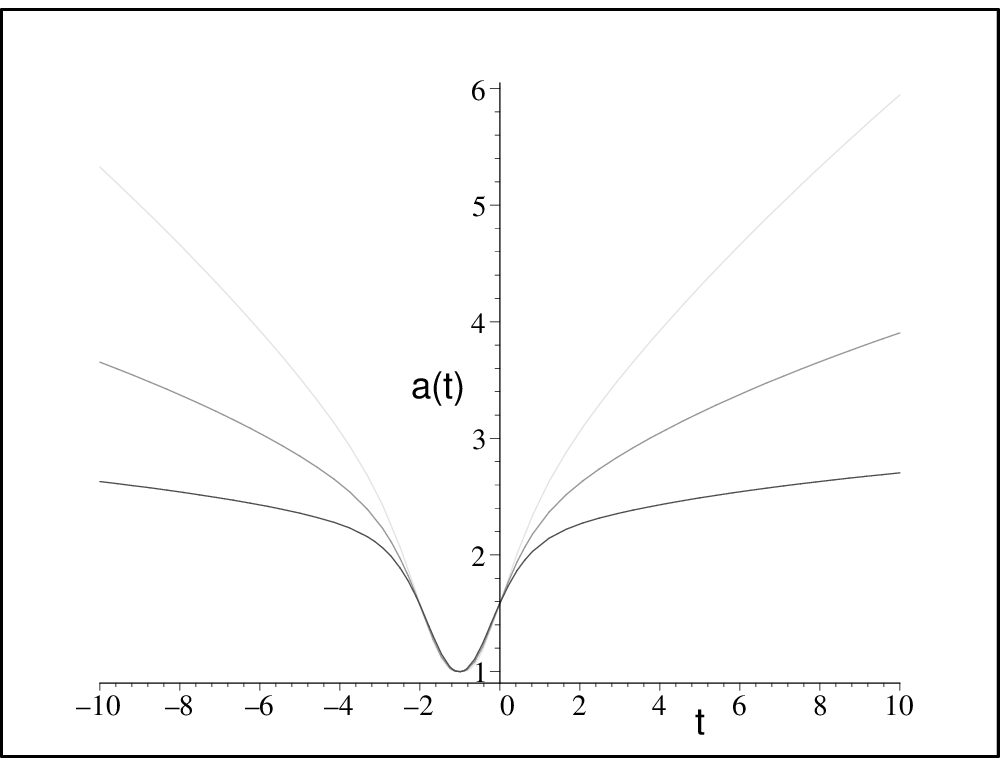, width = 10cm}\caption{
Examples of functions $a(t)$ that {\it would} be associated with
full-SD-branes with $k_{\parallel}=0$ and $k_{\perp}=-1$. From top
to bottom these correspond to $p=2$, $p=4$ and $p=6$.}
\label{consistent}}

At this stage we would like to reformulate the singularity theorem
of \cite{buchelwalcher} in a perhaps more geometric form that is
more closely related to a no-go theorem for full-SD-branes.
Depending on the intrinsic geometry of the SD-brane of interest,
{\it i.e.}, the value of $k_{\parallel}$ and $k_{\perp}$, the
constant of proportionality $v$ in (\ref{volume}) will vary but it
is irrelevant for the following analysis. Let us consider in
details the cases for which $k_{\parallel}=0$ and $k_{\perp}=-1$.
Asymptotically flat boundary conditions for a full-SD-brane are
then of the form
\begin{eqnarray}\label{boundary} \lim_{t\rightarrow \pm \infty}
a_{\parallel}(t) = a_{\pm
\infty}, \;\;\;\; \lim_{t\rightarrow \pm \infty}
\dot{a}_{\parallel}(t) = 0, \\ \nonumber \lim_{t\rightarrow \pm
\infty} a_{\perp}(t) = \pm (t + \kappa_{\pm \infty}), \;\;\;\;
\lim_{t\rightarrow \pm \infty} \dot{a}_{\perp}(t) = \pm 1,
\end{eqnarray}
where $a_{\pm \infty}$ and $\kappa_{\pm \infty}$ are non-zero
constants. We do not consider the boundary condition of the type
\beq \lim_{t\rightarrow \pm \infty} a_{\perp}= +t \eeq since this
will most likely lead to a curvature singularity as
$a_{\perp}\rightarrow 0$. The average scale factor $a(t)$
associated with the boundary conditions (\ref{boundary}) behaves
like \beq \lim_{t\rightarrow \pm \infty} a(t) = |t+k_{\pm
\infty}|^{\frac{8-p}{9}}. \eeq Figure \ref{consistent} shows, for
$p=2$, $p=4$ and $p=6$, the example of a functions $a(t)$ with
this asymptotic behavior. The region $t=0$ could either correspond
to the tachyon reaching the top of the potential (Type I
full-SD-branes) or to a turning point (Type II full-SD-branes).
For $t>0$ we have $\ddot{a}/a <0$, which is consistent with the
equations of motion, and for $t<0$ there is necessarily a region
for which $\ddot{a}/a>0$. Recall that the time-reversal symmetric
solutions were excluded from the start for consistency reasons.
The important point here is that a behavior as shown on figure
\ref{consistent} can only be obtained if there is a region in the
evolution of $a(t)$ such that $\ddot{a}/a>0$. In other words,
there is no solutions unless the volume of the full-SD-brane is
allowed to go through a phase of positive acceleration. We have
seen before that this cannot be accomplished in pure supergravity
and that adding the most obvious open string mode, the tachyon,
does not help for $p\leq 7$. Presumably, adding more exotic matter
or using a more general ansatz, as suggested in
\cite{buchelwalcher}, could be useful. The results in \cite{kmp}
suggest that allowing for sources inducing worldvolume
anisotropies is likely to provide positive results for
full-SD-branes.

\FIGURE{\epsfig{file=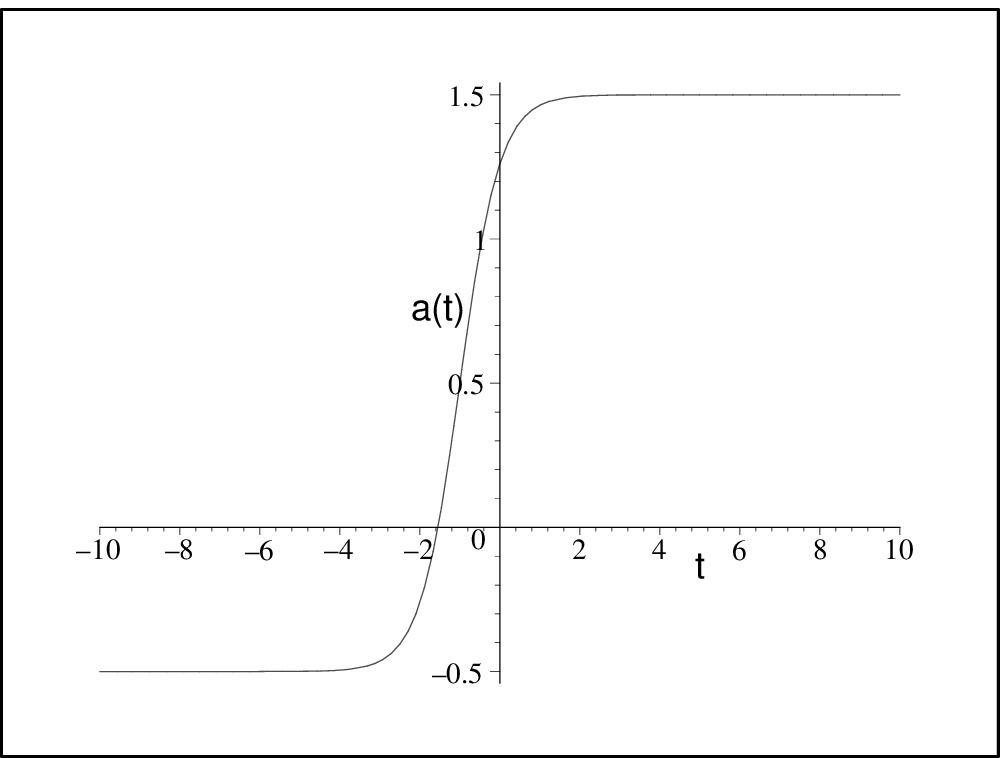, width = 10cm}\caption{
Example of a function $a(t)$ that {\it would} be associated with a
full-SD-brane with $k_{\parallel}=0$ and $k_{\perp}=0$.}
\label{consistent2}}

It therefore seems likely that full-SD-brane solutions with an
homogeneous and isotropic worldvolume simply do not exist in pure
supergravity and also when the extended sources are associated
with a tachyon with dynamics governed by a DBI-like action. This
conclusion is of course restricted to the specific ansatz for the
unstable branes density in the space transverse to their
worldvolume, namely, they were smeared along the corresponding
directions. It will be interesting to consider whether or not the
no-go theorem can be circumvented by using a less restrictive
ansatz\cite{four}. We should also note that other effective
actions for the tachyon dynamics are proposed in the literature
(see, {\it e.g.}, \cite{lambert,lambert2}). It would certainly be
interesting to check whether or not these actions would allow for
a period of positive acceleration which is a pre-requisite for
full-SD-brane solutions to exist.

As pointed out before, SD-branes can have different geometries
depending on the curvature of the worldvolume and of the
transverse space. For example for $k_{\parallel}=0$ and
$k_{\perp}=0$ (asymptotically flat) full-SD-branes are such that
\begin{eqnarray}\label{boundary2}
\lim_{t\rightarrow \pm \infty} a_{\parallel}(t) = {\rm const.},
\;\;\;\; \lim_{t\rightarrow \pm \infty} \dot{a}_{\parallel}(t) = 0, \\
\nonumber \lim_{t\rightarrow \pm \infty} a_{\perp}(t) = {\rm const.},
\;\;\;\; \lim_{t\rightarrow \pm \infty} \dot{a}_{\perp}(t) =
0. \end{eqnarray}
In this case a typical function $a(t)$ is shown on figure
\ref{consistent2}.  Again, this type of behavior can clearly not be
achieved unless there is matter in the system allowing a period of
constant positive acceleration for the overall volume of the geometry.

Geometric arguments of the type discussed here can clearly be
extended to all values of $k_{\parallel}$ and $k_{\perp}$. This
leads to a no-go theorem for homogeneous full-SD-branes in pure
supergravity and in the case where the supergravity modes
evolution is driven by an open string tachyon. Note that this is
only a reformulation of the singularity theorem of
\cite{buchelwalcher}. The latter simply says that that there will
always be a curvature singularity (or, more conservatively, a
region where $\dot{a}/a$ diverges) in the region $t<0$ of the type
of evolutions illustrated on figures \ref{consistent} and
\ref{consistent2}. The geometric no-go theorem was here
illustrated for Type I full-SD-branes which are those emerging
from a big-bang singularity. Type II full-SD-branes will roughly
be obtained by looking at the $t\rightarrow -t$ cases. The no-go
theorem applies to those as well.  Working from cosmological
intuition, we can also be pretty confident that the cases with
$k=+1$, for either the parallel or perpendicular scale factors,
will do a big-bang or big-crunch anyway, regardless of what type
of non-gravitational fields we include.

%--------------------------------------------------------------------+
\section{Space-filling SD-branes}

In the space-filling case, $p=8$, the conditions leading to the
singularity theorem of \cite{buchelwalcher} are no longer
satisfied. It is easy to see this by inspecting the form of
$I_{\rm tachyon}$ and $I_{\rm sugra}$ in eq.~(\ref{Itachyon2}).
%The energy density $\rho$ and pressure
%$P_\parallel$ of the tachyon are
%
%\beq\label{rhopee} \rho(T) = {\frac{\lambda V(T)
%e^{3\Phi/2}}{2\sqrt{\Delta}}} \, , \qquad P_\parallel(T) =
%-\rho(T) \Delta \eeq
%
%where, as usual, $\Delta\equiv 1-e^{-\Phi/2}{\dot{T}}^2$.  Of course,
%at large $t$ the dangerous-looking factor $1/\sqrt{\Delta}$ is
%ameliorated by the exponential damping from the potential.  The form
%of the potential is not crucial, although its large-$T$ behavior is
%known from string theoretic computations.  If needed, we can take
%$V(T)=1/\cosh(T/\sqrt{2})$.
%
For $p=8$ the tachyon contribution to the RHS of the equation for
${\ddot{a}}/a$ is
%
%\begin{eqnarray}
\beq\label{tgoo}
-{\frac{1}{9}} I_{\rm tachyon}
%=-{\frac{1}{9}} {\frac{1}{8}}\rho(T)[7-(p+1)\Delta]
%&=& {\frac{1}{9}}{\frac{1}{8}} {\frac{\lambda V(T) e^{\Phi(p/4-1/2)}
%}{2a_\perp^{8-p}\sqrt{\Delta}}} \left\{ (p-6)\left[ 1 -
%{\frac{(p+1)}{(p-6)}} {\dot{T}}^2 e^{-\Phi/2} \right] \right\}
%\nonumber \\
%&\stackrel{\rightarrow}{{\scriptscriptstyle
%p=8}}&
=+{\frac{1}{9\times 8}}{\frac{\lambda V(T) e^{3\Phi/2}
}{\sqrt{\Delta}}} \left[ 1 - {\frac{9}{2}} {\dot{T}}^2 e^{-\Phi/2}
\right].
%\end{eqnarray}
\eeq
At large times, this is negative (like the cases of larger
codimension which possess this property for all time).  However,
the $p=8$ tachyon contribution to the acceleration is {\em
positive} until $|\dot{T}| = e^{\Phi/4}\sqrt{2}/3$.

For the bulk\footnote{Of course, for $p=8$ we can drop the distinction
between $a$ and $a_\parallel$ since there is no transverse space.}
fields
\beq\label{Agoo} -{\frac{1}{9}}I_{\rm sugra} =
-{\frac{1}{2}}{\dot{\Phi}}^2 + {\frac{1}{16}} e^{-5\Phi/2}A^2 .
\eeq
Notice that the contribution of the Ramond-Ramond field is {\em
positive} definite.  The metric contribution is absent for $p=8$.

Therefore, the interesting question to ask is whether there always
exists some choice of initial conditions such that the big-crunch can
be avoided for a negative initial Hubble parameter.  (Note that we
expect no singularity trouble for positive initial Hubble parameter;
see next sections for details).

Let us consider the constraint carefully.  We have for
$p=8$, in Einstein frame,
\beq {\frac{8\times 9}{2}} \left[ H^2+{\frac{k^2}{a^2}} \right] =
\rho, \eeq
where the energy density is \beq \rho = {\frac{\lambda
V(T)e^{3\Phi/2}}{2\sqrt{\Delta}}} + \frac{1}{4}\dot{\Phi}^{2} +
\frac{1}{4}e^{-5\Phi/2}A^{2}. \eeq
Let us specify to the case $k=0$ for definiteness. Then we have for
$t=0$ the relation $H^2(t=0) = \rho(t=0)/36$.  So the size of the
initial Hubble parameter is set by
\beq\label{inigoo} \left| {\frac{{\dot{a}}(0)}{a(0)}} \right| =
{\frac{1}{6}} \sqrt{\rho(0)}. \eeq
The question is whether a small enough negative Hubble parameter can
be chosen at $t=0$, consistent with the constraint, such that the
positive acceleration at $t=0$ lifts it up to becoming a positive
Hubble parameter before the era of positive acceleration runs out.

The point to notice here is that the answer depends crucially on what
the Ramond-Ramond field is doing.  On the other hand, the dilaton
field cannot help sustain the period of positive acceleration, because
as we see above in eq. (\ref{Agoo}), it always contributes negatively
to the acceleration.  The dilaton also makes the era of positive
tachyon contribution to the acceleration shorter.  This is because
that era ends when $|{\dot{T}}|=e^{\Phi/4}\sqrt{2}/3$, and in a
typical half-SD-brane evolution, the string loop-counting parameter
$g_s e^\Phi$ falls.

If there are no Ramond-Ramond fields turned on at $t=0$, then it
looks rather unlikely to us that the SD8-brane could remain
nonsingular with a negative initial Hubble parameter.  Let us see
why, explicitly.  For the purposes of this argument we can turn
off the dilaton; as we saw above it only strengthens the
likelihood of developing a singularity. So for the moment we are
considering a situation with metric and tachyon only.  Now,
$\rho(0)$ sets the initial Hubble parameter, {\em and} the initial
positive acceleration, {\em and} whether or not SUGRA is actually
valid!  Explicitly, using eqs.~(\ref{tgoo}) and (\ref{Agoo}) we
have
\beq {\frac{{\ddot{a}}(0)}{a(0)}} = {\frac{1}{36}} \rho(T(0))
\qquad {\mbox{(for $g,T$ only)}}. \eeq
Here, we used the fact that the tachyon derivative is negligible at
$t=0$.

Now, in order for SUGRA to be a reasonable approximation to the
physics, we require small curvature.  Using eq.(\ref{ricciscalar}) for
the Ricci scalar at $t=0$ we find it to be of order $\rho(T(0))$ for
the tachyon contribution, of order $[{\dot{a}}(0)/a(0)]^2$ for the
gravity contribution, of order ${\dot{\Phi}}^2$ and of order
$e^{-5\Phi/2}A^2$ for the dilaton and Ramond-Ramond fields
respectively.  Therefore, each of these quantities, such as
$\rho(T(0))$, must be small.

Since for $\rho(0)$ small, $\sqrt{\rho(0)}$ is significantly
bigger, it looks unlikely that the era of positive tachyon
contribution lasts long enough to turn the Hubble parameter around
into positive territory.

On the other hand, this conclusion can change dramatically if we turn
on a nonzero Ramond-Ramond field at $t=0$, because in general
\beq {\frac{{\ddot{a}}(0)}{a(0)}} = {\frac{1}{36}} \rho(T(0))
\left[1-\frac{9}{2}{\dot{T}}^2(0)e^{-5\Phi(0)/2} \right] +
{\frac{1}{16}} e^{-5\Phi/2}A^2(0) - {\frac{1}{2}}{\dot{\Phi}}^2
\qquad {\mbox{(general)}}. \eeq
The Ramond-Ramond contribution to the positive acceleration has to
be small, in order for SUGRA to remain a decent approximation, but
it looks to us that it is quite possible for it to be big enough
to lengthen the positive-acceleration period.  In addition,
although the Ramond-Ramond contribution will typically fall with
time in an SD-brane evolution, we can see that it will persist in
contributing positively to the acceleration.

The equations are sufficiently complex, even in this restricted
ansatz of homogeneous tachyon\footnote{And, for $p<8$, branes
smeared in the transverse space}, that we cannot settle this
question absolutely definitively here.  However, it looks very
likely to us that it is {\em possible} for the SD8-brane to escape
the singularity theorem of \cite{buchelwalcher}.

One effect that various approaches to the problem of SD-brane SUGRA
fields has ignored thus far is particle production.  This will
typically slow down the rolling tachyon, and therefore potentially
prolong the period of positive acceleration.  This may help the
likelihood of finding a completely nonsingular SD8-brane.

We now move to the cases of interest for cosmology.

%--------------------------------------------------------------------+
\section{Cosmological applications}
\label{section:cosmo}

Cosmological scenarios involving tachyon condensation have been an
area of active investigation since the work of Sen and Gibbons
(see, {\it e.g.}, the review \cite{gibbonsreview}).  It is
particularly interesting to look at what the singularity theorem
of \cite{buchelwalcher} has to say about these situations, which
are of considerable interest to the early universe cosmology
community.  As we will show, the theorem of \cite{buchelwalcher}
simply does not apply; if it did, it would point to a quite
general inability to have nonsingular cosmologies for matter
satisfying physically reasonable energy conditions.  In
particular, an interesting {\em nonsingular} cosmology involving
plain ordinary $d=4$ Einstein gravity coupled to the tachyon
({\`{a}} la the DBI action) was actually first pointed out in
\cite{gibbons}.

Let us briefly recap the cosmology story of \cite{gibbons},
changing the action to general dimension $D$ (and renormalizing
the tachyon potential to conform to the conventions of
\cite{buchelwalcher,us2}).  We have
\beq S = {\frac{1}{16\pi G_{D+1}}} \int d^{D+1}x \sqrt{-g} \left[
{\cal R} - \lambda V(T)\sqrt{1-g^{\mu\nu}\nabla_\mu T \nabla_\nu
T} \right]. \eeq
The case of the homogeneous and isotropic rolling tachyon was
considered in \cite{gibbons}; subsequently the issues of
inhomogeneity were considered in \cite{kofmanetal}.

The metric ansatz for the homogeneous case can be taken to be
\beq ds^2 = -dt^2 + a^2(t) d{\vec{x}_D}^2 \,, \eeq
where $d{\vec{x}_D}^2$ is is either flat ($k=0$), spherical
($k=+1$) or hyperbolic ($k=-1$). The tachyon equation of motion is
not important for us here, except that it rolls fast (nearly at
${\dot{T}}=1$) at large $t$.

The constraint equation is
\beq \left({\frac{\dot{a}}{a}}\right)^2 + {\frac{k}{a^2}} =
{\frac{1}{D(D-1)}} {\frac{\lambda V(T)}{\sqrt{1-\dot{T}^2}}} \eeq
and can be considered as the initial condition at $t=0$.  For $k=0$,
$H={\dot{a}}/a$ must be nonzero at $t=0$ in order to satisfy the
constraint.  The cases $k=-1$ and $k=+1$ are presumably less
interesting for cosmology since there is convincing experimental
evidence that our universe is flat.  We can choose either the positive
or negative branch (for the initial Hubble parameter) for any given
cosmology.

The bulk equation of motion is
\beq {\frac{\ddot{a}}{a}} = {\frac{1}{D(D-1)}} {\frac{\lambda
V(T)}{\sqrt{1-\dot{T}^2}}} \left[ 1- {\frac{D}{2}}{\dot{T}}^2
\right]. \eeq
Notice that this is precisely the same result as we saw in the
previous section for the case $p=8$, if the dilaton is negligible.
This is of course a consistency check.

Following usual cosmological intuition, the {\em positive} Hubble
parameter branch solution of the Friedmann equation is chosen in
\cite{gibbons}.  Then, according to the bulk equation of motion
the scale factor starts out concave up, but as the tachyon gets
rolling fast it becomes concave down. But because the tachyon
obeys the weak energy condition ($\rho>0$), $\dot{a}/a$ {\em
always stays positive}. The long-time behaviour is
${\ddot{a}}/a\sim 0^-$, and $a(t)$ either approaches a {\em
constant} as $t\rightarrow\infty$ for $k =0$ or it is {\em linear
in time} for $k=-1$.  These give perfectly {\em nonsingular}
cosmologies.  For $k=+1$, however, there is a big-crunch.

These conclusions hold for any $D>2$, and in particular any dimension
$D$ that is sensible for cosmology.  All of this is very intuitive,
based on previous experience with ordinary minimally coupled scalar
fields in four-dimensional cosmology.

%- - - - - - - - - - - - - - - - - - - - - - - - - - - - - - - - - - +
\subsection{Comments on accelerating cosmology}

Although they are singular, the SD2- and SM2-brane spacetimes found in
\cite{andy1,gut,kmp} have been studied with the aim of finding
characteristics shared with our own universe \cite{acceleration}. The
corresponding time-dependent 10- and 11-dimensional spacetimes have
been shown to admit, for some values of the parameters in the
solutions, periods of positive worldvolume acceleration
\cite{acceleration}. There is convincing evidence that the universe is
currently in such an accelerating phase. Experimental data also
suggest that the dark energy component of the universe has
$\Omega_{D}\sim 70\%$. The papers \cite{acceleration} were considering
$\Omega_{D}=1$ which is clearly unrealistic. However progress towards
achieving more realistic models were made in \cite{gkl} where
additional external sources were added to the supergravity system.

Certainly, our approach consisting in adding unstable brane
sources follows the philosophy of \cite{gkl}. This approach is in
fact a more general study than what was done so far in tachyon
cosmology. It would be very interesting to see in details how
adding the dilaton and a Ramond-Ramond field would affect the
results obtained in tachyon cosmology in a more general way than
has been done thus far.

As a first baby step, we can ask whether the tachyon is likely to
prefer positive or negative worldvolume acceleration. It is clear
from the no-go theorem that \beq \label{uyi}
(p+1)\frac{\ddot{a}_{\parallel}}{a_{\parallel}} =
-(8-p)\frac{\ddot{a}_{\perp}}{a_{\perp}} - I_{{\rm sugra}} -
I_{{\rm tachyon}}. \eeq It was noted before that for $p=2$ we have
$I_{{\rm tachyon}}\geq 0$. Simply by inspection of eq.~(\ref{uyi})
we see that the tachyon contribution to the equations of motion
appears to favor negative rather than positive worldvolume
acceleration. Since $I_{{\rm sugra}}\geq 0$ the only way positive
worldvolume acceleration can be achieved is if
$\ddot{a}_\perp/a_\perp <0$ during the tachyon evolution.

%--------------------------------------------------------------------+
\section{Half-SD-branes}

In the supergravity approximation, the half-SD-branes, and the
unstable D-branes, are morally equivalent to this cosmological case of
\cite{gibbons} shown above.  In particular, small bulk kicks will be
needed at $t=0$, {\it i.e.}, we need at least one nonzero (small,
positive) Hubble parameter at $t=0$.  For the numerical solutions of
\cite{us2}, the {\em future} ($t\geq 0$) solutions are not ruled out
at all by the singularity theorem of \cite{buchelwalcher}.

Let us see this analytically.  In the smearing convention of
\cite{buchelwalcher}, the equations of \cite{us2} are easily
converted by replacing $\lambda\rightarrow\lambda/a_\perp^{8-p}$;
then we find that $a_{\parallel}(t)$ evolves ever upwards when it
starts at $t=0$ with a positive derivative; $a_\perp(t)$ evolves
ever upwards as well.
In particular, the geometric average scale factor ${{a}}(t)$
involved in the singularity theorem of \cite{buchelwalcher} starts
out with positive derivative, and it slows down with time (because
$\ddot{{a}}<0$), eventually asymptoting to a constant.  This is
clearly consistent with the asymptotics,
$a_\parallel(t\rightarrow\infty)\sim {\rm{const.}},\,
a_\perp(t\rightarrow\infty)\sim t$.  This $t\geq 0$ behaviour is
also thoroughly consistent intuitively with the results of
\cite{gibbons}. As a simple example let us consider the type of
evolutions illustrated on figures \ref{consistent} and
\ref{consistent2} where an half-SD-brane corresponds only to the
$t\geq 0$ part of the evolution: either the tachyon starts at the
top of the potential (perhaps with some initial kinetic energy
although the initial conditions $T(0)=0$ and $\dot{T}(0)=0$ are
consistent as long as there is kinetic energy in the bulk fields
at $t=0$) or the tachyon starts evolving away from the top of the
potential. The behaviour shown on the figures is such that
$\dot{a}/a\geq 0$ and $\ddot{a}/a\leq 0$ throughout the evolution.

Therefore, we conclude that the $t\geq 0$ parts of the spacetimes of
\cite{us2} were perfectly legitimate. Let us consider a particular
example, {\it i.e.}, that of an half-SD2-brane. We consider a solution
with the boundary conditions\footnote{The graphs are actually
displayed with metric components in string frame, for consistency with
our previous paper.  This is unimportant for present purposes, because
graphs in Einstein frame look similar.}
\begin{eqnarray}
\label{fullbc} T(0)=0 \, , \;\;\; \dot{T}(0)=0.1 \, , \;\;\;
a_{\parallel}(0)=0.1 \, , \;\;\; \dot{a}_{\parallel}(0)= 0.079 \,
, \;\;\; a_{\perp}(0)=1 \, , \nonumber
\\ \dot{a}_{\perp}(0) = 0 \, , \;\;\; \Phi(0)=-1 \, , \;\;\;
\dot{\Phi}(0) =0 \, , \;\;\; A(0) = 0 \, ,
\end{eqnarray} and the Ramond-Ramond field itself vanishes at $t=0$.
These initial conditions are consistent with the equations of
motion including the constraint.
Figures~\ref{1a}, \ref{3a}, \ref{2a}, \ref{4a} and \ref{5a}
respectively show the time evolution of the fields
$a_{\parallel}(t)$, $a_{\perp}(t)$, $e^{\Phi(t)}$, the
Ramond-Ramond field and $T(t)$. Figure~\ref{6a} shows the
acceleration of the SD-brane volume, {\it i.e.}, $\ddot{a}/a$.

\FIGURE{\epsfig{file=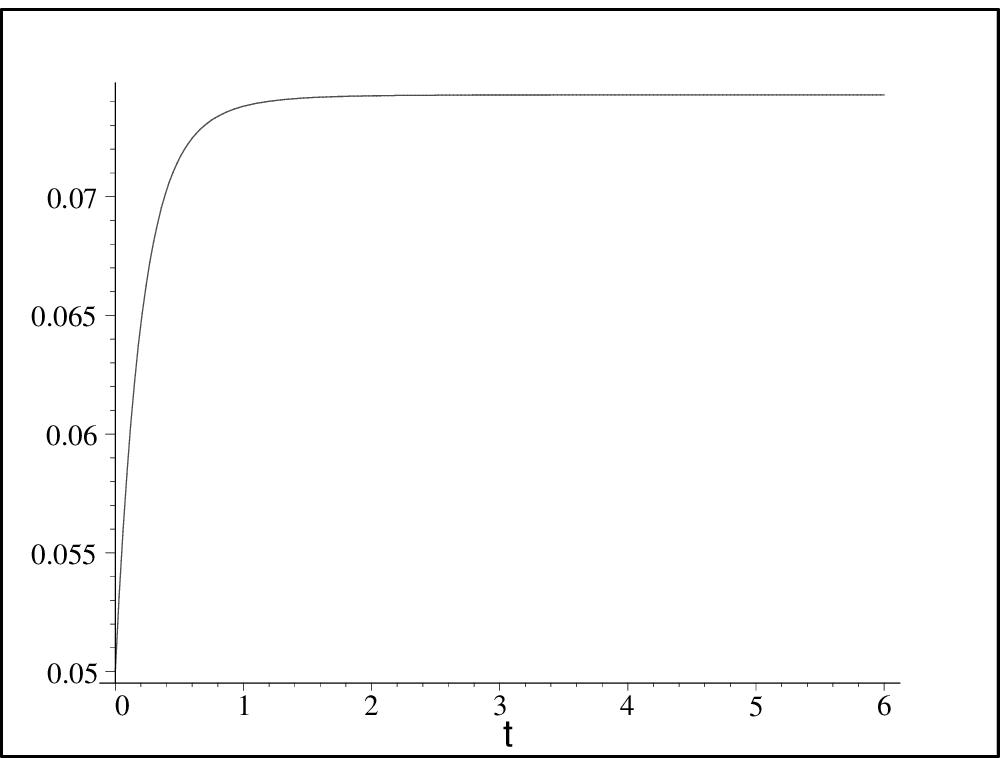, width = 10cm}\caption{The scale
factor on the worldvolume of a supergravity half-SD2-brane with
boundary conditions (\ref{fullbc}).} \label{1a}}

\FIGURE{\epsfig{file=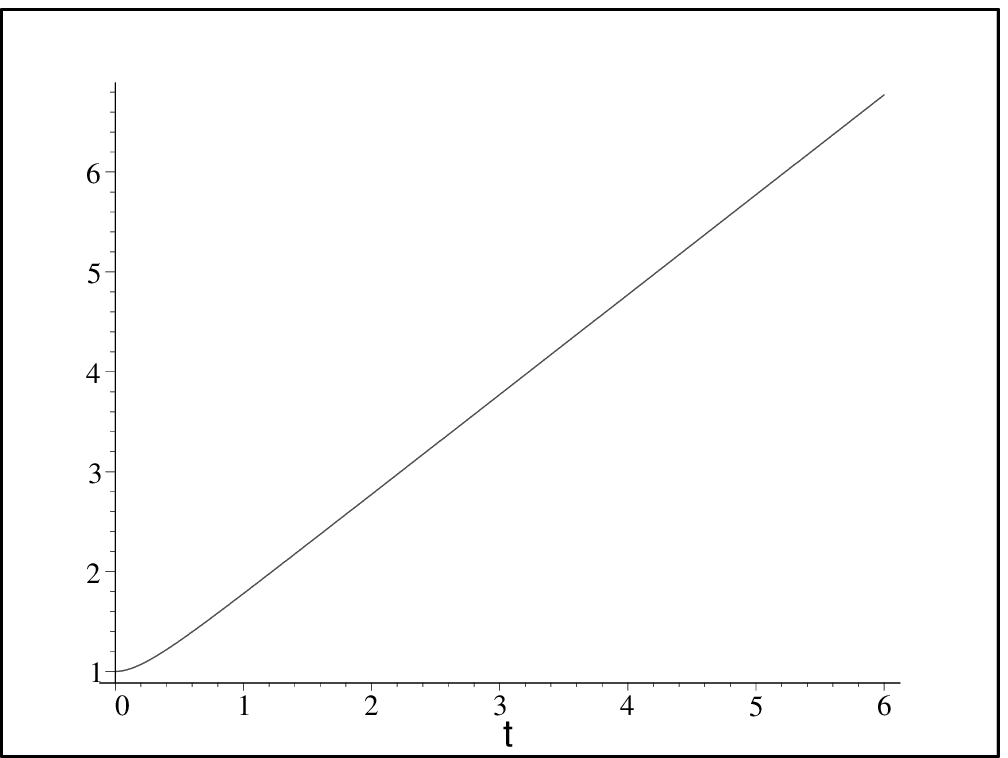, width = 10cm}\caption{The transverse
scale factor $a_{\perp}(t)$ for a supergravity half-SD2-brane with
boundary conditions (\ref{fullbc}).} \label{3a}}

\FIGURE{\epsfig{file=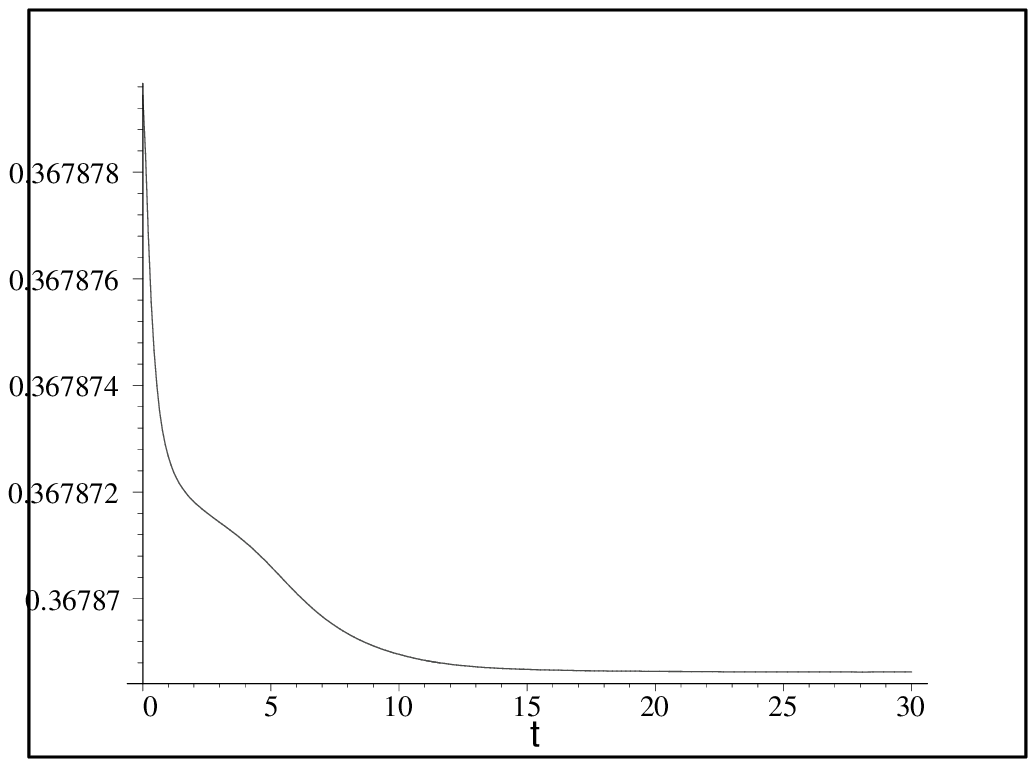, width = 10cm}\caption{The
dilaton function $e^{\Phi(t)}$ for a supergravity half-SD2-brane
with boundary conditions (\ref{fullbc}).} \label{2a}}

\FIGURE{\epsfig{file=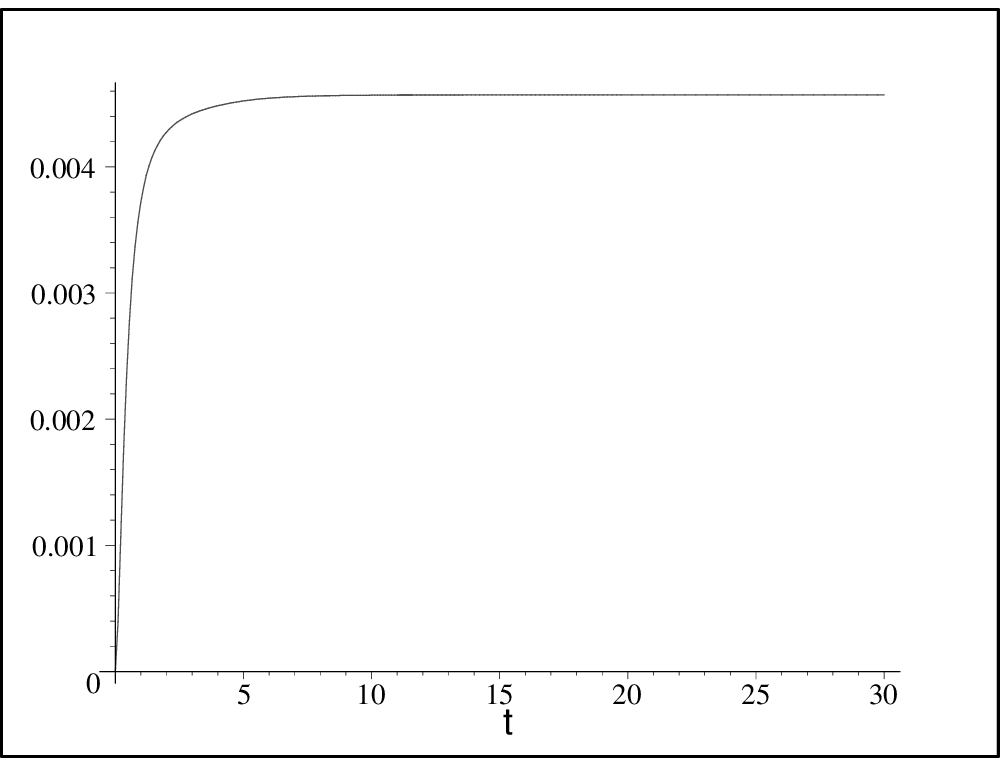, width = 10cm}\caption{The
Ramond-Ramond field for a supergravity half-SD2-brane with
boundary conditions (\ref{fullbc}).} \label{4a}}

\FIGURE{\epsfig{file=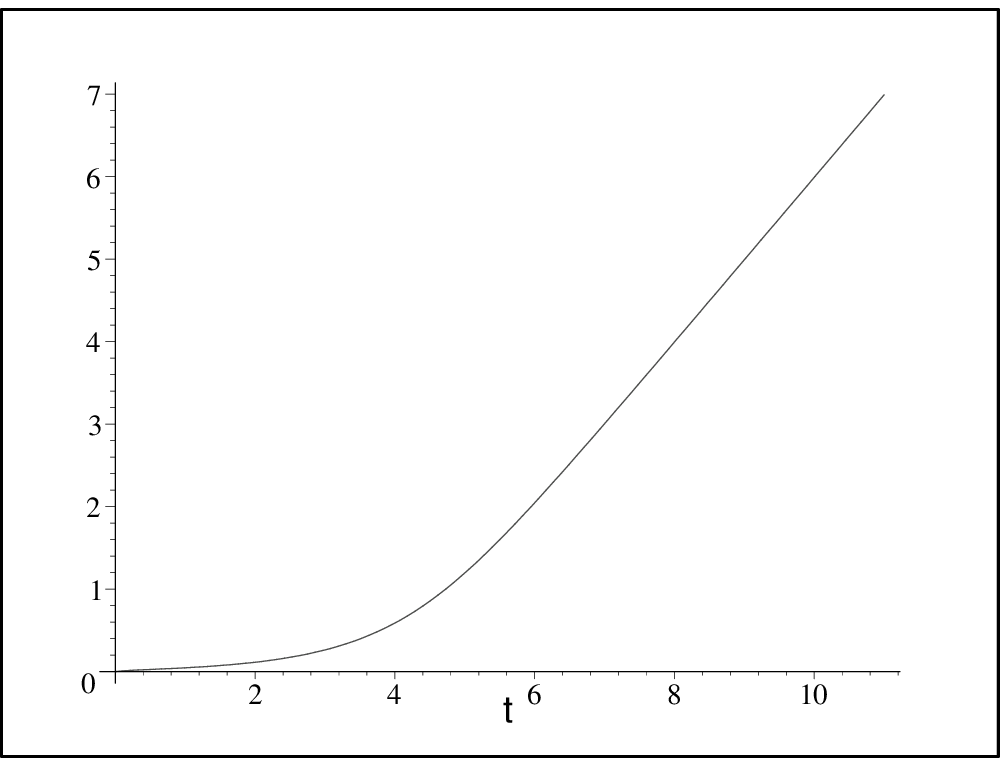, width = 10cm}\caption{The open
string tachyon field associated with a supergravity half-SD2-brane
with boundary conditions (\ref{fullbc}).} \label{5a}}

\FIGURE{\epsfig{file=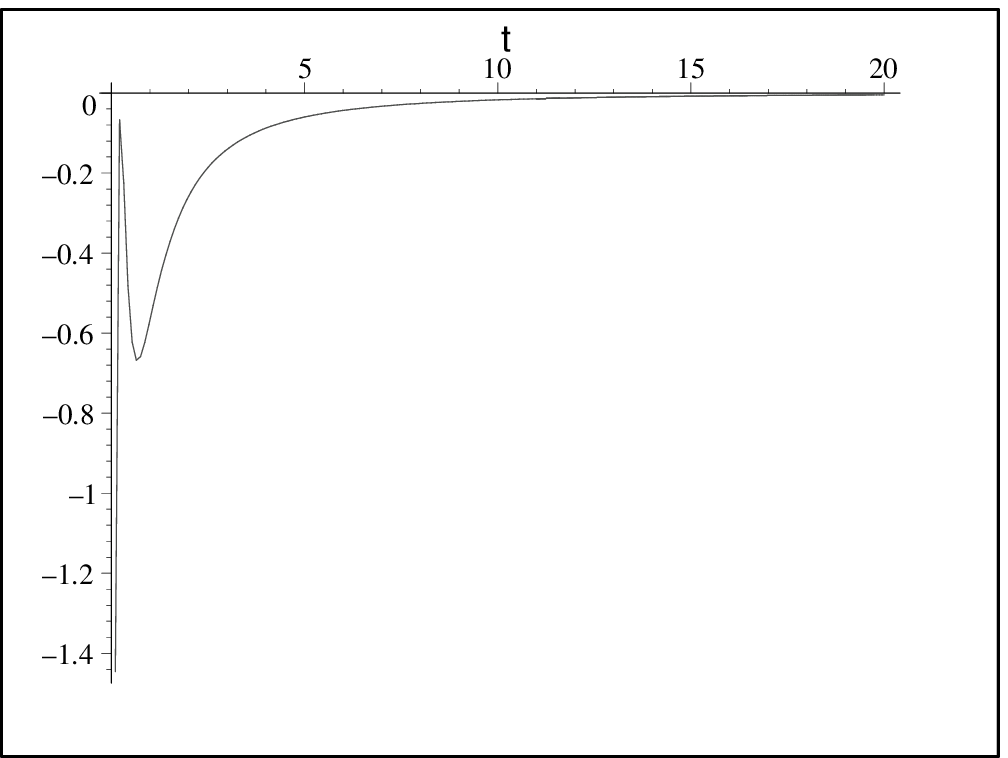, width = 10cm}\caption{The
acceleration $\ddot{a}/a$ for the half-SD2-brane with boundary
conditions (\ref{fullbc}).} \label{6a}}

%%%
The singularity theorem of \cite{buchelwalcher} therefore shows
that the numerical code of \cite{us2} must have been truly
unstable - rather than simply exhibiting the flat-space
tachyon-matter numerical instability - on the {\em past} side of
the SD-brane evolution. This potential problem was addressed in
section 4.4.1 of \cite{us2} but evidence at the time suggested it
was only an artefact of the numerical analysis.
An easy way to see this one-sidedness analytically is to consider
the easy trick used in \cite{us2} to actually evolve numerically
from near the tachyon hilltop to $t\rightarrow-\infty$: we swap
the signs of time and time derivatives of bulk fields at $t=0$
(and, depending on whether we are considering same-side or
opposite-side full SD-branes, we swap the sign of $f(T)$ as well,
or not).  This means that we are starting the evolution with {\em
negative} Hubble parameter, and since $\ddot{{a}}/a$ is still
negative even with sign-swapped time, we must reach a big-crunch
after a finite time in the past of $t=0$.  This conclusion of a
past singularity applies whether or not we consider ``same-side''
or ``opposite-side'' full SD-branes, {\it i.e.}, whether the
tachyon rolls back to the same minimum or the other minimum.
%%%

%--------------------------------------------------------------------+
\section{Discussion}

In this short note, we have shown in precise detail with the aid of
the new analytic singularity theorem of \cite{buchelwalcher} exactly
where the numerical results of our previous work \cite{us2} broke
down.  We have also pointed out that cosmology with rolling tachyons
is of course not ruled out by the singularity theorem. In particular,
half-SD-branes certainly exist as well-behaved supergravity solutions,
provided that small (positive) bulk kicks are given to at least one
bulk field at $t=0$, such as $a_\parallel(t)$.  As we saw in the
cosmology section \ref{section:cosmo}, this initial condition choice
is in tune with textbook cosmological intuition.

Most interesting directions for future work are manyfold.  We have
already pointed out here (in section 2) that anisotropy on the
worldvolume may provide a way of escaping the singularity theorem of
\cite{buchelwalcher}.

The general question of inhomogeneities in tachyon cosmology is
currently under further investigation \cite{inhomogtree} as well.
In particular, in the current context it will be interesting to
know whether the singularity theorem of \cite{buchelwalcher} can
be extended to cover those cases.  Intuition based on standard
cosmology indicates that inhomogeneity will probably only
strengthen the no-go theorem, but this must be checked, since
inhomogeneities notably complicate all the field equations.

Another thing to check will be the role of {\em un}smearing the
branes in the transverse space.  This is important because the
validity of the no-go theorem is limited by the restrictive nature
of the SUGRA ansatz used in deriving it.  We feel this issue has
particular importance because we have pointed out that the
SD8-brane may actually avoid the no-go theorem.  And the $p=8$
case is really the only one ``honestly'' covered by the ansatz
used.

It could be interesting to include in this investigation also the
question of whether changes in the low-energy effective action for
the tachyon, as suggested in \cite{lambert}, may change the
outcome.

The details of how the tachyon rolls can be significantly affected
by particle production when backreaction is included. In
particular, if the rolling slows down because of particle
production, there may be a longer time of positive acceleration
for the space-filling case.

Finally, there is the question of whether quantum string theoretic
ingredients are unavoidably necessary for description of
SD-branes. In \cite{buchelwalcher} particular emphasis was put on
the fact that the nonexistence of SD-brane supergravity solutions
- established via the no-go theorem - is in tune with their
picture that SD-brane decay should be essentially
quantum-mechanical.  We think that this issue depends crucially on
whether the space-filling SD-branes escape the no-go theorem or
not.  If they do, then it would seem rather odd that whether or
not quantum stringy ingredients are needed depends so crucially on
the codimension.

It is possible that quantum stringy ingredients may turn out to be
important for getting long periods of positive acceleration in
models of this type. Clearly, an important issue is whether SUGRA
can remain a decent approximation to the physics in such a
situation.  Let us suppose for the sake of argument that this can
be done.
The program of \cite{blw,us2} consists in solving the Einstein
equations with sources in the form of a perfect fluid with
stress-energy tensor components
\beq \label{s1} \rho = \frac{\lambda
V(T)e^{\phi(p/4)-1/2}}{2a_{\perp}^{8-p}\sqrt{\Delta}} +
\frac{1}{4}\dot{\Phi}^{2} + \frac{1}{4}e^{a\Phi}A^{2},  \eeq
\beq \label{s2} P_{\parallel} = -\frac{\lambda
V(T)e^{\phi(p/4)-1/2}\sqrt{\Delta}}{2a_{\perp}^{8-p}} +
\frac{1}{4}\dot{\Phi}^{2} - \frac{1}{4}e^{a\Phi}A^{2}, \eeq \beq
\label{s3} P_{\perp} = \frac{1}{4}\dot{\Phi}^{2} +
\frac{1}{4}e^{a\Phi}A^{2}. \eeq
Each of these sources separately satisfies
the dominant energy condition (see
\cite{ellis} for definitions). Of course, this implies that the
weak energy condition and the null convergence condition are
satisfied as well. It was noted in \cite{buchelwalcher} that the
strong energy condition,
\beq  7 \rho + (p+1)P_{\parallel} + (8-p)P_{\perp} \geq 0, \eeq
is satisfied by eqs.~(\ref{s1})--(\ref{s3}) for $p<7$. As shown in
section \ref{nogot}, circumventing the no-go theorem would, as a
minimum requirement, require that a period of positive
acceleration (for the volume $V_{{\rm S}}$ of the spacetime) be
allowed. This could be achieved by introducing new terms ({\it
i.e.}, sources with a different equation of state) to the RHS of
(\ref{bcond}) that contribute positively, {\it i.e.}, in such a
way as to violate the strong energy condition. Sources in the form
of massive open or closed strings might achieve the desired
behavior therefore resolving the singularity and circumventing the
no-go theorem. Analysis of massive open string mode production
during tachyon evolution were performed in
\cite{andythermo,stro2,stro1}. BCFT calculations for the emission
of massive closed strings can be found in
\cite{other2,maldacena2}.

To illustrate our point let us use a simple example, {\it i.e.},
that of a scalar field in four dimensions, with the usual with
stress-energy tensor. An implication of the strong energy
condition being satisfied is that
\beq
\left(T_{ab}-\frac{1}{2}g_{ab}T^{c}_{c}\right)\xi^{a}\xi^{b}\geq
0, \eeq
which becomes
\beq \label{jk} \left(\nabla_{a}\phi\, \xi^{a}\right)^{2}
-\frac{1}{2}m^{2}\phi^{2} \geq 0 \eeq
for a massive scalar field. $\xi^{a}$ is a timelike unit vector.
After using the equation of motion and integrating over a region
$M$ of the spacetime the LHS of (\ref{jk}) becomes
\beq \label{violate} \frac{1}{2}\int_{M}\left( g^{ab} + 2
\xi^{a}\xi^{b} \right)\nabla_{a}\phi\nabla_{b}\phi \, d^{4}x -
\frac{1}{2}\int_{\partial M} \phi \nabla_{a}g^{ab}d n_{b}, \eeq
where $\partial M$ is a boundary and $d n_{b}$ a normal vector.
The first term in eq.~(\ref{violate}) is always positive while the
second term, which is negative-definite, will remain small as long
as the region of integration $M$ is large compare to the Compton
wavelength ($\lambda=1/m$) of the mode. The strong energy
condition might therefore be violated in small regions of the
spacetime in particular when the domain of existence (in time and
space) of the modes is small. This simple line of reasoning
suggest that the emission of massive modes (open and/or closed)
might provide a loophole to the singularity theorem for
full-S-branes. Of course,
%awp new
as we mentioned above,
this assumes that the tachyon evolution, including emission of
massless and massive open and closed strings is smooth and that
the contribution of the massive modes does not overwhelm that of
the massless modes.  Consideration of this issue is also under
investigation.

%--------------------------------------------------------------------+
\section*{Acknowledgements}

AWP would like to acknowledge useful discussions with Andrei
Frolov and Lev Kofman.  Both authors would also like to thank Alex
Buchel and Johannes Walcher for generously sharing a draft of
their singularity theorem paper before it appeared on hep-th.

FL was supported in part by NSERC of Canada and FCAR du Qu\'ebec.
Research of AWP is supported by the Canadian Institute for Advanced
Research (CIAR), NSERC of Canada, and the Alfred P. Sloan Foundation
of the USA.

\bigskip
%--------------------------------------------------------------------+

%--------------------------------------------------------------------+
\end{document}